# Analysis of the railway heave induced by soil swelling at a site in southern France

Anh-Minh Tang[1], Yu-Jun Cui[1], Viet-Nam Trinh[1], Yahel Szerman[2], Gilles Marchadier[2*]

[1] *Université Paris-Est, UR Navier-CERMES, Ecole des Ponts*
[2] SNCF, Direction de l'Ingénierie

**Corresponding author**

Prof. Yu-Jun CUI
Ecole Nationale des Ponts et Chaussées, CERMES
6-8 Avenue Blaise Pascal, Cité Descartes, Champs-sur-Marne
F-77455 MARNE-LA-VALLEE CEDEX 2
France

Email: cui@cermes.enpc.fr
Phone: +33 1 64 15 35 50
Fax: +33 1 64 15 35 62

[*] This paper is dedicated to the last co-author who died on May 29, 2007, at the age of 30.

**Abstract**: In order to better understand the heave observed on the railway roadbed of the French high-speed train (TGV) at Chabrillan in southern France, the swelling behaviour of the involved expansive clayey marl taken from the site by coring was investigated. The aim the study is to analyse the part of heave induced by the soil swelling. First, the swell potential was determined by flooding the soil specimen in an oedometer under its in-situ overburden stress. On the other hand, in order to assess the swell induced by the excavation undertaken during the construction of the railway, a second method was applied. The soil was first loaded to its in situ overburden stress existing before the excavation. It was then flooded and unloaded to its current overburden stress (after the excavation). The swell induced by this unloading was considered. Finally, the experimental results obtained were analyzed, together with the results from other laboratory tests performed previously and the data collected from the field monitoring. This study allowed estimating the heave induced by soil swelling. Subsequently, the part of heave due to landslide could be estimated which corresponds to the difference between the monitored heave and the swelling heave.

## 1. Introduction

Structural and pavement damage due to expansive clays have been observed in numerous countries as Saudi Arabia (Al-Shamrani and Dhowian, 2003; Abduljauwad et al., 1998; Al-Mhaidib, 1999), Australia (Fityus et al., 2004), Poland (Kaczynski and Grabowska-Olszewska, 1997), Turkey (Erguler and Ulysay, 2002), Oman (Al-Rawas and Qamaruddin, 1998), China (Shi et al. 2002), etc. Very often, the heave induced by soil swelling are predicted using numerical models through the analysis of soil moisture and volume changes in expansive soils (Vu and Fredlund, 2004; Masia et al., 2004; Wray et al., 2005). These models in general require a big number of parameters, some of which are difficult to be determined using common geotechnical tests. In this regard, empirical methods present their obvious advantage: the swell potential is estimated indirectly from the soil physical and geotechnical properties using empirical equations (Yilmaz, 2006; Abduljauwad and Al-Sulaimani, 1993). The swell potential is simply defined as the maximal swell percent obtained when the soil is flood by water under constant stresses.

The swell behaviour of soils is usually investigated in laboratory condition. Several methods have been elaborated permitting to determine the swell pressure and swell potential (Basma et al., 1995; Al-Shamrani and Dhowian, 2003; Al-Mhaidib, 1999). It has been shown that the experimental method employed could affect the test results. El-Sohby and Rabba (1981) noted that the swell potential of remoulded clayey soils depends on the initial water content, the type of clay mineral, the initial dry density, the clay content and the type of coarse grained fraction. Based on the laboratory test results, several methods have been elaborated and validated using data from field monitoring. Yoshida et al. (1983) proposed a method based on the general theory for unsaturated soils. Data from the monitoring of the movements of a floor slab activated by leakage of a water pipe buried under the floor slab was compared with the calculation. The comparison showed that this method, which required the data from oedometer test as well as the initial water content profile, gave satisfactory calculation results. Abduljauwad et al. (1998) observed the heave induced by artificially wetting the soil ground through vertical sand drains and compared it with the results from conventional oedometer tests. They concluded that the oedometer tests over-estimated the swell potential because of the rigidity of the oedometer ring. Al-Shamrani and Dhowian (2003) investigated the relevance of the swell parameters, obtained under oedometer and triaxial conditions, in predicting the potential heave of expansive soils. The results obtained from laboratory tests were analyzed based on the data from field testing. These authors also noted that oedometer tests over-estimated the in-situ heave and by contrast, tests in triaxial condition gave better estimation.

In the present paper, observations of the damage and the heave recorded on the roadbed of a high-speed railway line at Chabrillan, southern France, were first presented. This was mainly attributed to two mechanisms: (i) swelling of the marls beneath the railway roadbed; and (ii) landslide towards the railway lines. The aim of the present study is to investigate the heave induced by the soil swelling. A specific laboratory testing program was carried out on core samples and the swell potential was determined using two oedometer testing procedures. The analysis on the laboratory results and the field observation allowed assessing the part of the observed heave due to the involved marl swelling. Moreover, the effect of ground excavation was clearly evidenced, showing the importance to consider the unloading-swelling coupling in the heave analysis.

## 2. History of the site studied

The "LGV Méditerranée" is a French high-speed railway line connecting Saint-Marcel-lès-Valence and Marseille. The construction of this 250-km line started in 1996 and the line has been exploited commercially since June 2001. The Chabrillan site is located between Km 529+500 and Km 530+510 and the earthwork of the excavation was completed in July 1998. Figure 1 presents the elevation of the natural land before the excavation and the elevation of the line after the excavation for the location from Km 530+000 to Km 530+500. The depth of the excavation, the difference between the two elevations, was significant and varied from 9 to 34 m.

After the detection of the damages on a road near the Chabrillan excavation in 2001, a study was conducted by the French National Railway Company (SNCF) and the French Railway Network Management Company (RFF), in order to understand the causes of the damages. This study, which included field observations and numerical simulation, confirmed a landslide towards the railway line with a velocity of 1 mm/days in the location between Km 500+280 and Km 500+580. A second excavation of 600 000 $m^3$ was then undertaken on the top of the slope during 35 days (from April 9 to May 14, 2001) in order to neutralize this landslide. A plan view of the Chabrillan excavations and a representative cross-section are presented in Figure 2. More details about the study on the landslide as well as the excavation in 2001 can be found in Montagnon (2001) and Pouitout et al. (2001). Since April 2001, the Chabrillan site has been under surveillance by topographical monitoring (SNCF, 2007).

Owing to the intensification of damages observed on the ditches besides the railway line, a technical visit was organized in April 2005 and some pictures taken during this visit are shown in Figure 3. Figure 3*a* shows that the bottom of the ditch was damaged by the heave of soil; the retaining wall at the hill side was broken (Figure 3*b*); and the ditch was dislocated (Figure 3*c*). In order to identify the effect of soil swelling on the heave, which has been observed by topographical monitoring and was visible to the naked eyes, several boreholes were rotary-drilled in 2002 and 2007. The locations of these boreholes are shown in Figure 2; six boreholes (B1-1 to B1-6) on Line 1 to Paris and three boreholes (B2-1 to B2-3) on Line 2 to Marseille. Geotechnical characterizations have been performed on the soil taken from these boreholes in order to investigate its swell potential.

## 3. Description of the site, geological observations and geotechnical properties of investigated soils

The Chabrillan site is located at 30 km in the north-east of Montélimar (04° 45' 06" E, 44° 33' 34" N), a city of southern France in the Drôme department. The annual rainfall at Montélimar is 913 mm/year and the sunshine duration is 2498 h/year. The air temperature generally varies from about – 15°C to 40°C. In Figure 4, the monthly rainfall recorded at Montélimar from January 2003 to March 2007 is shown (after http://meteociel.fr). The data show that the monthly rainfall was generally strong (more than 100 mm) during the three last months of each year and it was generally low (less than 50 mm) during the summer (from June to August).

The first series of boreholes was drilled in February 2002. During this campaign, six boreholes (B1-1, B2-1, B2-2, B1-4, B1-5 and B2-3) were drilled on the two sides of the railway line (see Figure 2) at 4-5 m depth. Intact soil specimens taken from these boreholes were used for the determination of geotechnical parameters and for performing oedometer swelling tests (Fondasol, 2002). A second series of boreholes (B1-2, B1-3, and B1-6) was drilled in January 2007. These three boreholes were deeper: 8-12 m. Intact soil specimens taken from the boreholes were used for oedometer swelling tests in the present work.

The profiles of the soil observed in the boreholes are presented in Figure 5. The first layer (A: Clayey sand and gravel), having a thickness of 0.90 – 2.50 m, corresponds to the structure of the roadbed. Below this layer, there are green clayey marl (B), grey calcareous marl (C) and grey clayey marl (D). The layers E and G involve siltstone and marly limestone, respectively. A layer of white and rose limestone (F) was found at deeper level (below 5 m). The first observations showed that the grey calcareous marl (C) and the limestone (G and F) are as hard as mudstone. As a consequence, oedometer swell tests were not able to be carried out on these soils. The siltstone (E) does not contain any clay fraction and was thus considered insensitive to water. Only green clayey marl (B) and grey clayey marl (D) were then investigated to study the swelling behaviour from oedometer tests.

The involved calcareous soils was analysed by Poitout et al. (2001) and it was found that these soils are in fact marls or calcareous marls with high carbonate contents (% $CaCO_3$ = 45 – 90%). They have variable water content (w = 3 – 14%) and show high uniaxial compression strengths (8 – 17 MPa). The clayey marls contain a significant carbonates proportion (% $CaCO_3$ = 28 – 40%) and they are fine-grained soils (% < 80 μm = 100%; % < 2 μm = 40 - 75%). Table 1 presents the results from laboratory identification tests (after Fondasol, 2002). For the soil B (B1-5, 2.42 – 2.53 depth), the liquid limit $w_L$ = 46%, the plasticity index $I_P$ = 15%, and the bleu methylene value (VBs = 3.68) are higher than other samples which correspond to soil C. The characteristics of the clayey marls in the present work are similar to that studied in the works of Derriche and Cheikh-Lounis (2004), Windal et al. (2002), Paaza et al. (1998) and Aiban (1995).

## *4. Field observations*

The altimetric monitoring of the railway roadbed has been undertaken by levelling (i) the rivets embedded in the legs of the catenary poles, and (ii) the metal rods embedded in the concrete railway sleepers that situate between the catenary poles. Figure 6 presents the heave observed on the legs of the catenary poles of Line 1 from April 2001. In Figure 7, the results of Figure 6 are presented as a function of the distance from the Km 530-000 for every year. That allows observing the evolution of the profile of soil displacement in Line 1 with time. Levelling observed on the legs of the catenary poles of Line 2 started in November 2003 and the results are presented in Figure 8. Levelling of the metal rods embedded in the concrete railway sleepers started in April 2001, and the results are presented in Figure 9 for Line 1 and in Figure 10 for Line 2.

Observations on the heave of the railway roadbed show that the heave can reach 80 mm for the six-year period from April 2001 to March 2007. In the case of Line 1 (Figure 6, Figure 7 and Figure 9), the heave observed in the sections from Km 530-067 to Km 530-132 and from Km 530-267 to Km 530-458 varied in the range of 30 to 80 mm. In addition, the heave observed in these sections showed increasing trends. For others sections, the observed heave was less than 20 mm for a six-year period and has almost reached the stabilisation since December 2003. In the case of Line 2, the most significant heaves (higher than 20 mm for the period from April 2001 to March 2007) were observed in the sections between Km 530-267 and Km 530-438 (Figure 10). The heaves in these sections presented increasing trend while the heaves in other sections have stabilised since December 2003. In addition, Figure 8 shows that the heaves observed during the four-year period, from November 2003 to March 2007, at Km 530-285, Km 530-290 and Km 530-335 were higher than 18 mm.

On the other hand, all the results of heave monitoring show a fluctuation of about ± 3 mm. This fluctuation can be attributed to the effect of seasonal climatic changes on expansive soils: (i) during the winter where rainfall was important, water infiltration was higher than evaporation and the soil trended to swell; (ii) by contrast, during the summer, evaporation quantity exceeded water infiltration, the soil trended to shrink.

The topographical monitoring of the railway roadbed has been performed, from a polygonal network of control points constituted of rivets embedded in the railway sleepers. The displacements in the direction perpendicular to the railway lines were calculated from this monitoring data and from the theoretical location of the lines. The results are presented in Figure 11 where positive values correspond to the displacements in the direction from the excavation performed in 2003 towards the lines axis (see the cross-section A-A in Figure 2). From April 2001 to June 2001, the lateral displacement was measured every week. The results presented in Figure 11 show a fluctuation of ±7 mm during this short period. This fluctuation can be attributed to the accuracy of this method in which the location of the points was calculated indirectly. In spite of this, significant displacements (larger than 20 mm) in the direction perpendicular to the lines axis were identified for the points at Km 530-107, Km 530-334, and Km 530-413 for the period from April 2001 to March 2007.

## *5. Oedometer swelling tests*

The soil cores taken from the boreholes (90 mm diameter) were inserted in plastic tubes having ends closed and transported to the laboratory for testing. That allowed protecting the soil water content during the transport as well as the storage in laboratory. In order to perform oedometer swelling test, 20-mm long section was cut at the location to be tested using a saw for metal. The confining ring of the oedometer having a sharp edge was then pushed inside the soil sample. The surfaces of the soil specimen were finished using a sharp steel straight edge. The final dimensions of the soil specimen were 70 mm in diameter and 10 mm high. The confining ring having the soil specimen inside was then installed in the oedometer cell.

The "loaded-swell method" (after Al-Rawas et al. 1998) was first applied to evaluate the swell potential of the soil. Following this method, the soil was first loaded at its natural water content to the current field overburden stress. After that, the soil was flooded and allowed to swell, under the constant applied overburden stress, until swelling ceases. This vertical swell was recorded and considered as the swell potential.

The details of the tests performed following the loaded-swell method are presented in Table2. In this table, the tests SC1-6 were performed in 2002 (Fondasol, 2002) and the tests LS01-06 were performed in 2007. In the present work, six tests were performed on soil taken from two boreholes (B1-2 and B1-6) at various depths. These specimens correspond to the clayey marls identified previously (see Figure 5). The initial dry density of the soil is equal to the dry mass (soil specimen taken after the test and oven-dried at 105 °C for 24 h) divided by the initial volume of the specimen (70 mm in diameter, 10 mm high, that corresponds to a volume of 38 465 mm$^3$). The initial and final water contents ($w_i$, $w_f$) were calculated from the difference between the mass of soil sample before and after the test and the dry mass. The overburden stress of the soil ($\sigma_v$) was estimated using the following equation:

$\sigma_v = z\gamma + \sigma_0$ [1]

where: $z$ is the depth of the soil specimen; $\gamma$ is the mean unit weight of the soil $\gamma = 20$ kN/m$^3$; and $\sigma_0$ is the stress induces by the structure of the railway roadbed, $\sigma_0 = 30$ kPa. The stress applied in the test was rounded from the estimated overburden stress.

The results of the tests performed following this method are presented in Figure 12 where the changes in height and the vertical strains are plotted versus time. For all the six tests, an initial settlement was observed during the application of the field overburden stress. The settlement was considered to be stabilized when the rate of settlement was less than 0.01 mm/8 h following the French Standard (AFNOR, 1995). Water was added after the stabilization of the settlement (at t = 24 h for LS04 and LS06; t = 42 h for LS01 and LS03; and t = 70 h for LS02 and LS05). For all the six tests, the water flooding gave rise to soil swelling.

The settlements due to loading and the swells due to water flooding are summarised in Table2. For the tests performed in 2007, the settlements varied from 0.200 to 0.458 mm, corresponding to a vertical strain of 2.00 - 4.58%. The swells were comprised between 0.016 and 0.030 mm (0.16 and 0.30%). Similar procedure was applied to determine the swell potential of soil taken from the first series of boreholes which was drilled in 2002 (Fondasol, 2002). The swell measured upon flooding under the overburden stress was: 0.30 – 0.80%. It can be noted that the swell potential determined in 2002 was higher than that determined in 2007. The water content was determined before and after the tests. All the tests showed that the final water content $w_f$ was higher than the initial one $w_i$. That means the soil absorbed water during the test.

In order to evaluate the swell mobilised by the excavation undertaken in 1998 during the construction of the railway line, a second method was proposed. The soil was first loaded at its natural moisture content to its overburden stress existing before the excavation in 1998. After the stabilisation of the settlement, the soil was flooded and allowed to swell under this overburden stress until swelling ceases. Finally, the soil was unloaded to the current overburden stress (after the excavation in 1998). The overburden stress existing before excavation in 1998 was estimated using the following equation:

$\sigma_v = (z + h)\gamma$                         [2]

where $z$ and $\gamma$ are similar to that used in equation [1], $h$ is the depth of the excavation undertaken in 1998 (see Figure 1 and Table 3). This method was then called "loaded-unloaded-swell method".

Three tests were carried out. The tests conditions as well as the samples initial and final states are summarized in Table 3. The obtained results are presented in Figure 13 where the changes in height and vertical strains are plotted versus time. It can be observed that the initial loading to the overburden stress existing before the excavation in 1998 induced a large settlement in all the tests. Flooding under this high value of stress (at t =100 h for test LUS01; t = 94 h for test LUS02; and t = 75 h for test LUS03) did not induce significant swell. By contrast, unloading to the existing overburden stress (after excavation in 1998) induced a significant swell (at t = 450 h for test LUS01; t = 138 h for test LUS02; and t = 94 h for test LUS03).

It can be observed in Table 3 that the initial settlement due to loading under the overburden stresses existing before excavation in 1998 ($P_1$) was very large; it varied from 0.464 to 1.136 mm (corresponding to a vertical strain of 4.64 and 11.36%, respectively). The swell induced by flooding ($Sw1$) was very low; it varied from 0.000 to 0.023 mm (corresponding to 0.00 to 0.23%, respectively). However, the swell induced by unloading ($Sw2$) was high; it varied from 0.065 to 0.130 mm (corresponding to 0.65 to 1.30% of swelling strain respectively).

## *6. Analysis*

The heave observed on the railway roadbed during a six-year period (Figures 7 – 10) can be attributed to two different mechanisms: (i) swelling of the marls beneath the railway roadbed; and (ii) landslide towards the railway lines. In the present paper, only the swelling of the marls beneath the railway roadbed is analysed. The swelling mechanism can be explained by the schematic view presented in Figure 14 where the soil volume (*v*) is plotted versus time *t* (Figure 14*a*) and versus the overburden stress $\log\sigma_v$ (Figure 14*b*). In Figure 14*a*, the time $t_0$ corresponds to the beginning of the excavation in 1998. The initial state of the soil before this excavation is represented by the point A in the *v*- $\log\sigma_v$ plot (Figure 14*b*). The excavation (path A-B) induced an immediate swell of soil. After that, the soil continued to absorb water and swell (path B-C-E-F). After Chao et al. (2006), the time required for the soil to reach its maximal swell depend strongly on the soil hydraulic properties, the hydrological conditions of the site, and the thickness of the soil layer. This value may vary from a few minutes (in laboratory scale) to several years (in field scale).

Oedometer is usually used to investigate the settlement and the heave behaviour of soil. One of the advantages of this test is that the soil stress and strain conditions are close to the "in-situ" conditions: zero lateral strain and controlled vertical stress. Obviously, the oedometer test can not exactly simulate the "in-situ" conditions which are rather two-dimension even three-dimension problem. It can be considered as a simple method to analyze the soil swelling behaviour in the field. The results of the test LS02 (load-swell method) are also presented in Figure 14*c,d* and that of the test LUS02 (load-unload-swell method) in Figure 14*e,f*.

The moment $t_1$ corresponds to the sampling of the drilled cores in 2002. In the *v*-$\log\sigma_v$ plot, this sampling corresponds to an unloading path (path C-D). During the loaded-swell tests, the specimens were reloaded to its overburden stress (path D-C) and then flooded to swell (path C-E-F). When performing triaxial tests on soil specimens taken from great depth, Delage et al. (2007), Graham et al. (1987) applied the "in-situ" stress prior to re-saturate the soil. This procedure avoids the large soil swelling which may induce a microstructure change, resulting in a modification in mechanical behaviour. In fact, extracting the soil from depth correspond to stress release. It can be then expected that imposing the "in-situ" stress allow the sample to recover "in-situ" state. This assumption is valid only when the natural water content of soil was well conserved during the storage of soil samples. The swell potential obtained from the loaded-swell tests performed in 2002 correspond to the swell $s_{t1}$ in Figure 14*a*. In the same fashion, the time $t_2$ corresponds to the sampling of drilled cores performed in 2007 (path E-G). During the loaded-swell tests performed in 2007, the soil followed the path G-E-F. And the swell potential obtained (path E-F) corresponds to the swell $s_{t2}$. (Figure 14*c,d*).

The loaded-unloaded-swell method was applied in order to evaluate the total swell potential induced by the excavation performed in 1998 ($s_{t0}$). In fact, from the drilled cores sampled in 2007, the soil specimens were loaded to the overburden stress that existed before the excavation (path G-E-A). Then, after flooding under this stress, the soil was unloaded to the current overburden stress (path A-B) and let swell under inundated conditions (path B-C-E-F). Figure 14*e, f* present the results of the test LUS02. The total swell obtained from this unloading path (path A-B-C-E-F) corresponds then to the swell $s_{t0}$ in Figure 14*a*.

Following these interpretations, the swell potential of the soil determined from the oedometer swelling tests is plotted versus time in Figure 15: (i) the swell potential in 1998 was determined by the loaded-unloaded-swell method (represented by *%Sw2* in Table 3); (ii) the swell percent in 2007 was determined by the loaded-swell method (represented by swell potential in Table2); (3) and the swell potential in 2002 was also determined by the loaded-swell method (Fondasol, 2002). Significant data scatter was observed. This could be attributed to the natural soils heterogeneity, samples disturbance during sampling and transport, etc. Nevertheless, it can be observed that the swell potential of soil has been decreasing with time. That confirms the explication above on the swelling mechanism (Figure 14).

For further analysis, the total heaves and the lateral displacement (for the period from April 200 to March 2007) are plotted versus the distance from the location Km 530-000 in Figure 16. It can be observed that the heaves measured on Line 1 (Figure 16*a*) reached its maximal values ($\cong$ 80 mm) at Km 530-100 and Km 530-350; on Line 2 (Figure 16*b*), heave reached its maximal value ($\cong$ 60 mm) at Km 530-440. The lateral displacement observed on Line 1 (Figure 16*c*) was much lower than the heave observed (less than 30 mm).

The information about the profile of soil is equally plotted in Figure 16. From the profiles of the boreholes (Figure 5), the total thickness of the layers including Green clayey marl (soil B) and Gray clayey marl (soil D) was calculated for each borehole and considered as the total thickness of clayey marl layers. In Figure 16, these thicknesses are plotted versus the location of the boreholes. As shown in Figure 5, only three boreholes (B1-2, B1-3, and B1-6) were drilled deeper than 5 m although clayey marls were observed at even 9-m depth (B1-2). For the analysis shown in Figure 16, the total thickness of clayey marls was calculated for 5-m depth and 10-m depth separately. In the case of Line 1 (Figure 16*a*), by considering only the layer of soil from 0 to 5-m depth, the total thickness of clayey marls layers have its highest values at Km 530-100 (1.40 m) and Km 530-375 (2.55 m). These locations correspond to the highest heave observed ($\cong$ 80 mm). Similar remarks can be made when considering the layer from 0 to 10 m depth. In the case of Line 2 (Figure 16*b*), the total thickness of clayey marls layers is highest at Km 530-430 (0.95m), this location corresponds equally to the highest heave observed on Line 2 ($\cong$ 60 mm).

Following the results shown in Figure 15, the swell potential decreased from 0.35-0.90% (in 2002) to 0.15-0.30% (in 2007). That means a swell percent of 0.05-0.75% would take place for the period from 2001 to 2007. By multiplying this swell percent to the total thickness of clayey marls layers for the depth up to 10 m (Figure 16*a*), the heave induced by soil swelling for the period from April 2001 to March 2007 can be estimated: (1) At Km 530-110, 6.3-m thickness corresponds to a swell of 3 - 47 mm; (2) At Km 530-180, 2.1-m thickness corresponds to a swell of 1 - 16 mm; (3) At Km 530-375, 3.1-m thickness corresponds to a swell of 2 - 23 mm. It can be noted that the estimated heave is lower than the heave recorded during this period (Figure 16*a*) at Km 530-110 and Km 530-375 . The estimated heave at Km 530-180 (1 – 16 mm) is similar to the recorded one (between 7 mm at Km 530-177 and 12 mm at Km 530-197). In addition, following the estimation shown in Figure 15, the soil would continue to swell 0.15 – 0.3% for the next ten years.

## 7. Discussion

The fluctuation of the monitored heave (± 3 mm) at the Chabrillan site can be likely attributed to the annual climate changes. Similar fluctuations of ground movement were recorded at Maryland field site near Newcastle, Australia (after Fityus et al., 2004; Masia et al., 2004) and at two sites in Texas (Wray et al., 2005). These works confirmed that soil suction decreases during winter, leading to soil swelling; and soil suction increases during summer leading to soil shrinkage. The fluctuation of the recorded ground movement was ± 20 mm and the suction changes were observed at a depth up to 2 m. The fluctuation of ground movement recorded in the present work is much lower (± 3 mm). In fact, a layer of clayey sand and gravel, 0.90 – 2.50 m thick, covers the ground (see Figure 5) and that would minimize the effect of climatic changes on the suction changes of clayey marls and subsequently on the ground movement.

The soil swelling mechanism showed in Figure 14 was validated by the results obtained from oedometer swell tests (Figure 15) where a clear decrease of swell potential with time was observed. Abduljauwad et al. (1998) studied the swelling behaviour of an expansive clay (hydraulic conductivity $k = 10^{-10}$ m/s) and observed that the swell stabilised after 1 day for a conventional oedometer test on a sample of 10-mm high while it took 14 months for a laboratory slab test on a sample of 250-mm high. That can be explained by the water infiltration in unsaturated soils: after Yong and Mohamed (1992), when wetting an unsaturated soil column, the distance between the wet front and the water source can be described as a linear function with the square root of time. That means the time needed to saturate a column is $n^2$ times longer with a column $n$ times higher. In the work of Abduljauwad et al. (1998), the slab test had a sample 25 times higher than the oedometer sample. The time needed to reach the equilibrium in the oedometer test was 1 day. The time estimated to reach the equilibrium in the slab test is then 25×25 = 625 days (21 months). This value is in the same order of magnitude of the value observed experimentally (14 months). In the present work, a few hours were required for the stabilisation of swell in oedometer tests (see Figure 12 and Figure 13), that is in agreement with the work of Al-Rawas et al. (1998). On the other hand, following the results shown in Figure 15, approximately 20 years are required in field conditions to reach the stabilisation of the swell mobilised by the excavation performed in 1998.

The results obtained from oedometer swelling tests following the two adopted procedures have allowed estimating the evolution of the swell potential with time. Subsequently, the heave induced by soil swelling for the period from April 2001 to March 2007 has been estimated. It was found that the heave induced by soil swelling was lower than the heave monitored in the field for this period at the locations having thick clayey marl layers (Km 530 -110 and Km 530 375). Obviously, this difference would be attributed to the heave induced by the landslide that was evidenced by the lateral displacement monitoring (Figure 16*c*) on one hand and, on the other hand to the swell of other soils that have not been taken into account in the study. For instance, the green calcareous marl (soil C in Figure 5) is as hard as a mudstone and it was not able to be hand-trimmed for performing oedometer tests. For this reason, the swell potential of this soil was ignored. It is however believed that the swelling potential of these hard soils would be relatively limited.

Al-Rawas (1999) carried out a laboratory testing program on undisturbed expansive soils and rocks. Two of the conclusions drawn from the investigation were: (1) the use of index properties such as liquid limit, plasticity index and activity for quantifying the potential heave of expansive materials can be misleading; (2) performing detailed engineering tests is the most reliable method for assessing the actual swelling potential. The analysis done in the present work can be used to explain these conclusions drawn by Al-Rawas (1999). In fact, as shown in Figure 14 and Figure 15, the swell potential of the soil decreases with time. That explains why the use of index properties can not quantify the potential heave of expansive soil. On the other hand, to predict the heave induced by soil swelling, it is important to identify the mechanism that mobilized the soil swelling: the seasonal changes can induce soil swell or shrinkage and that can be identified by annual fluctuation of ground movement (Fityus et al. 2004); heave can be induced by significant changes in the hydrological conditions of the site as a leakage in water line buried under the floor slab (Yoshida et al. 1983), excavation unloading (Mersi et al. 1994), injection of water in the field (Abduljauwad et al. 1998), etc.

## *8. Conclusion*

Heave-induced damages were observed at the site of Chabrillan in southern France where a deep excavation (9 – 34 m depth) was carried out during the construction of a high-speed railway line in 1998. Two mechanisms have been supposed to be the causes of the heave: landslide towards the line and swelling of the clayey marls beneath the line. The aim of the paper is to analyse the heave induced by soil swelling. This swelling was supposed to be mobilised by the excavation performed during the construction of the line. In fact, analysis of field data showed that the ground movement due to climatic changes was limited to ± 3 mm.

Oedometer swelling tests were performed following two methods: loaded-swell method and loaded-unloaded-swell method. The first method allowed determining the swell potential of the soil at the moment of sampling (in 2002 by Fondasol and 2007 in the present work). The second method allowed estimating the total swell originated by the excavation performed during the construction of the line (in 1998). Only green clayey marl (B) and grey clayey marl (D) were investigated, other soils having been too hard to be tested.

The results showed a clear decrease of the soil potential with time. Extrapolation of the relationship between swell potential and time allowed estimating the time required in field conditions to reach the stabilisation: about 20 years from the excavation in 1998. As far as the heave is concerned, a good agreement between the total thickness of clayey marls layers and the ground heave was identified, confirming the contribution of these clayey marls to the monitored heave. In addition, it has been found that the part induced by the swelling of clayey marls was in general lower than the monitored heave, showing that the landslide evidenced by the lateral displacement monitoring was also an important cause for the heave. Note however that the swell of other soils has been ignored in the study.

## *Acknowledgement*

The French National Railway Company (SNCF) and the French Railway Network Management Company (RFF) are gratefully acknowledged for providing field monitoring data and for the financial support.

**Table 1. Geotechnical characterization of some soil specimens (After Fondasol, 2002)**

| Borehole | Depth (m) | Initial $w$ (%) | $w_L$ (%) | $I_p$ | VBs |
|---|---|---|---|---|---|
| B1-1 | 2.38 – 2.50 | 17.1 | 42 | 15 | 2.57 |
| B1-4 | 2.70 – 2.98 | 10.1 | 37 | 13 | 2.41 |
| B1-5 | 2.42 – 2.53 | 21.8 | 46 | 15 | 3.68 |
| B2-1 | 4.45 – 4.83 | 17.4 | 42 | 18 | 2.70 |
| B2-2 | 2.47 – 2.59 | 18.5 | 43 | 18 | 2.38 |
| B2-3 | 4.75 – 4.85 | 11.8 | 43 | 19 | 3.51 |

**Table2. Results obtained from the tests following the loaded-swell method.**

| Test | BH | year | z (m) | $\rho_i$ (Mg/m$^3$) | $w_i$ (%) | $w_f$ (%) | $\sigma_v$ (kPa) | Sett. (mm) | Sw (mm) | %Sw (%) |
|------|------|------|-----------|---------------------|-----------|-----------|------------------|------------|---------|---------|
| SC1  | B1-1 | 2002 | 1.8 - 3.05 | 1.92 | 13.6 | 14.4 | 67 | – | – | 0.40 |
| SC2  | B1-4 | 2002 | 1.7 - 2.8  | 2.00 | 11.9 | 13.2 | 84 | – | – | 0.50 |
| SC3  | B1-5 | 2002 | 1.6 - 2.8  | 1.67 | 21.8 | 23.2 | 56 | – | – | 0.80 |
| SC4  | B2-1 | 2002 | 3.95 - 5.1 | 1.69 | 19.1 | 21.1 | 168 | – | – | 0.30 |
| SC6  | B2-3 | 2002 | 3.9 - 4.95 | 2.09 | 10.0 | 10.4 | 168 | – | – | 0.30 |
| LS01 | B1-2 | 2007 | 4.40 | 2.11 | 11.28 | 12.70 | 125 | 0.342 | 0.016 | 0.16 |
| LS02 | B1-2 | 2007 | 5.02 | 2.04 | 13.55 | 14.39 | 150 | 0.200 | 0.020 | 0.20 |
| LS03 | B1-2 | 2007 | 7.80 | 1.85 | 18.15 | 19.82 | 200 | 0.365 | 0.021 | 0.21 |
| LS04 | B1-6 | 2007 | 2.10 | 1.72 | 23.16 | 25.58 | 75  | 0.272 | 0.027 | 0.27 |
| LS05 | B1-6 | 2007 | 2.30 | 1.62 | 24.03 | 26.22 | 75  | 0.353 | 0.030 | 0.30 |
| LS06 | B1-6 | 2007 | 7.74 | 1.64 | 23.10 | 23.54 | 175 | 0.458 | 0.027 | 0.27 |

*BH: Borehole; z: depth of the soil specimen; $\rho_i$: initial dry density; $w_i$ : initial moisture content; $w_f$ : final moisture content; P: vertical stress applied; Sett.: settlement due to the loading of P; Sw: swell induced by flooding at P; %Sw: swelling strain*

**Table 3. Results obtained from the tests following the loaded-unloaded-swell method**

| Test | BH | h (m) | z (m) | $\rho_i$ (Mg/m$^3$) | $w_i$ (%) | $w_f$ (%) | $P_1$ (kPa) | Sett. (mm) | Sw1 (mm) | $P_2$ (kPa) | Sw2 (mm) | %Sw2 (%) |
|---|---|---|---|---|---|---|---|---|---|---|---|---|
| LUS01 | B1-6 | 20 | 2.55 | 1.66 | 21.17 | 21.59 | 450 | 0.792 | 0.012 | 100 | 0.093 | 0.93 |
| LUS02 | B1-6 | 20 | 7.76 | 1.49 | 28.52 | 29.34 | 550 | 1.136 | 0.000 | 175 | 0.130 | 1.30 |
| LUS03 | B1-2 | 10 | 5.00 | 1.87 | 17.14 | 17.58 | 300 | 0.464 | 0.023 | 150 | 0.065 | 0.65 |

*BH: Borehole; h: Depth of the excavation undertaken in 1998; z: depth of the soil specimen; $\rho_i$: initial dry density; $w_i$: initial moisture content; $w_f$: final moisture content; $P_1$: overburden stress before excavation in 1998; Sett.: settlement due to the loading of $P_1$; Sw1: swell induced by flooding at $P_1$; $P_2$: overburden stress after excavation in 1998; Sw2: swell induced by unloading from $P_1$ to $P_2$; %Sw2: swelling strain induced by unloading from $P_1$ to $P_2$;*

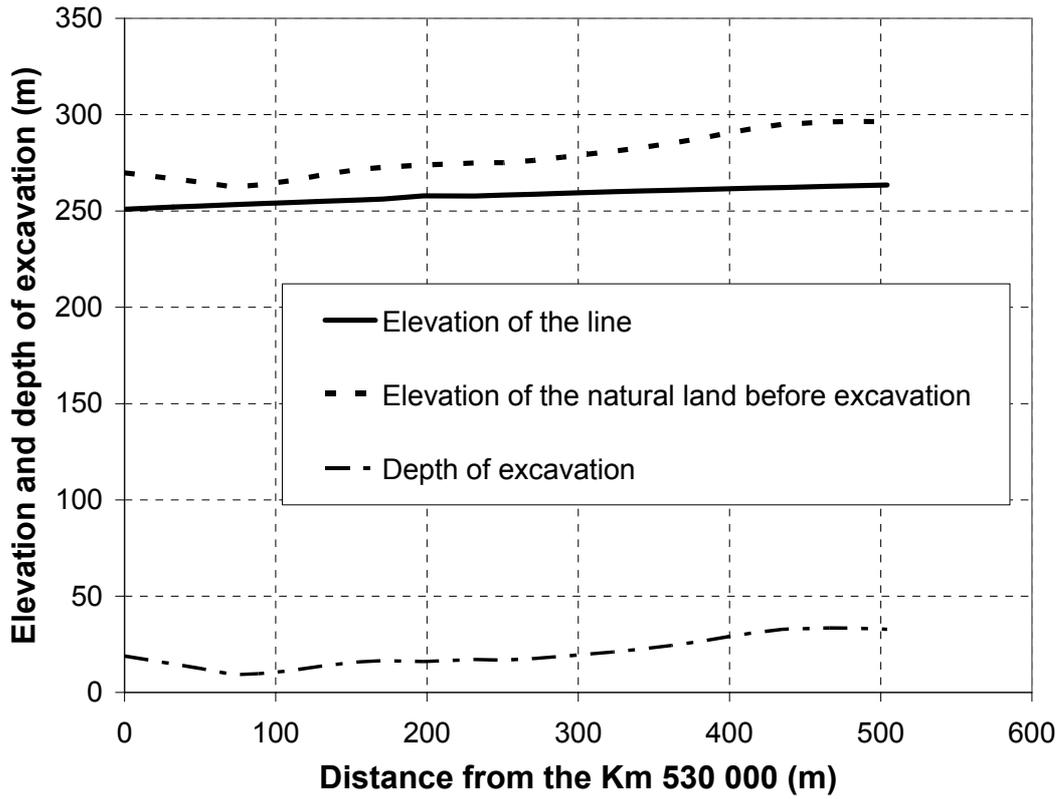

**Figure 1.** Elevations of the natural land (before excavation in 1998) and of the line (after excavation in 1998).

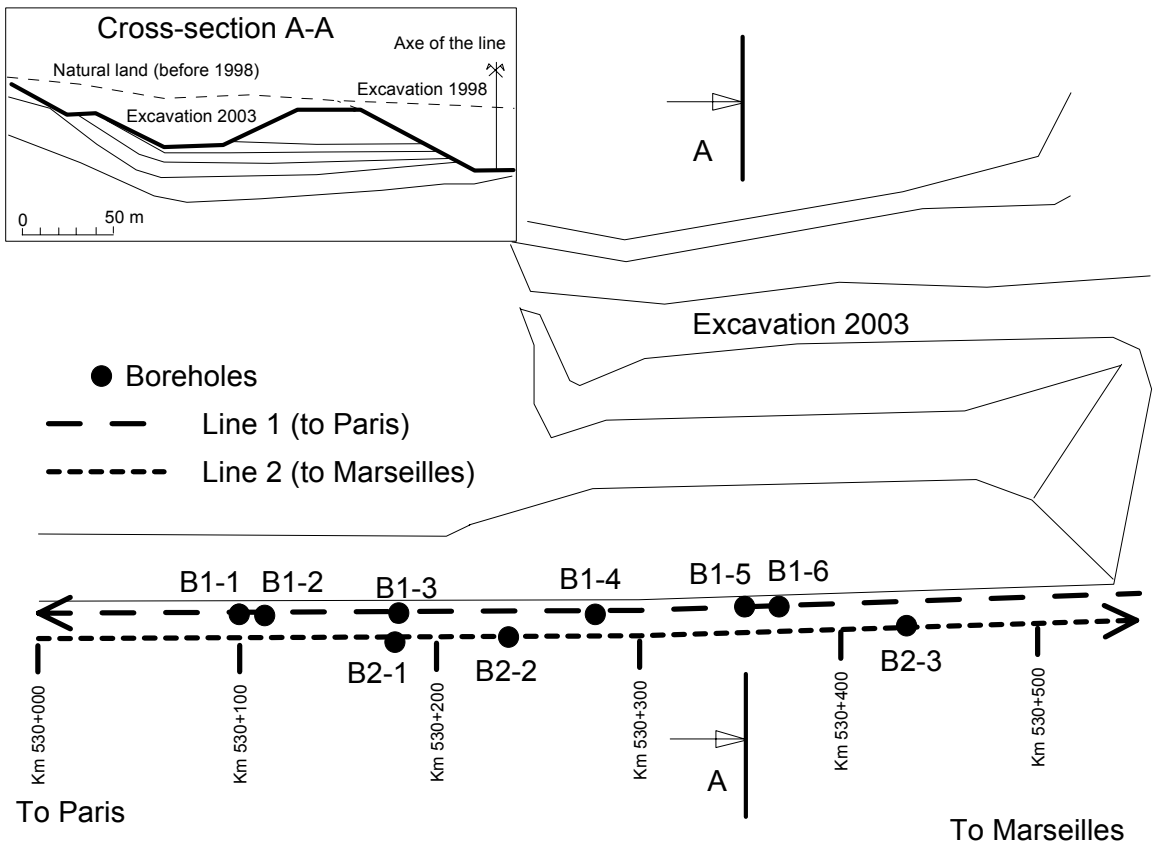

**Figure 2. Plan view of Chabrillan excavations and a representative cross-section. (Modified from Poitout et al. 2003).**

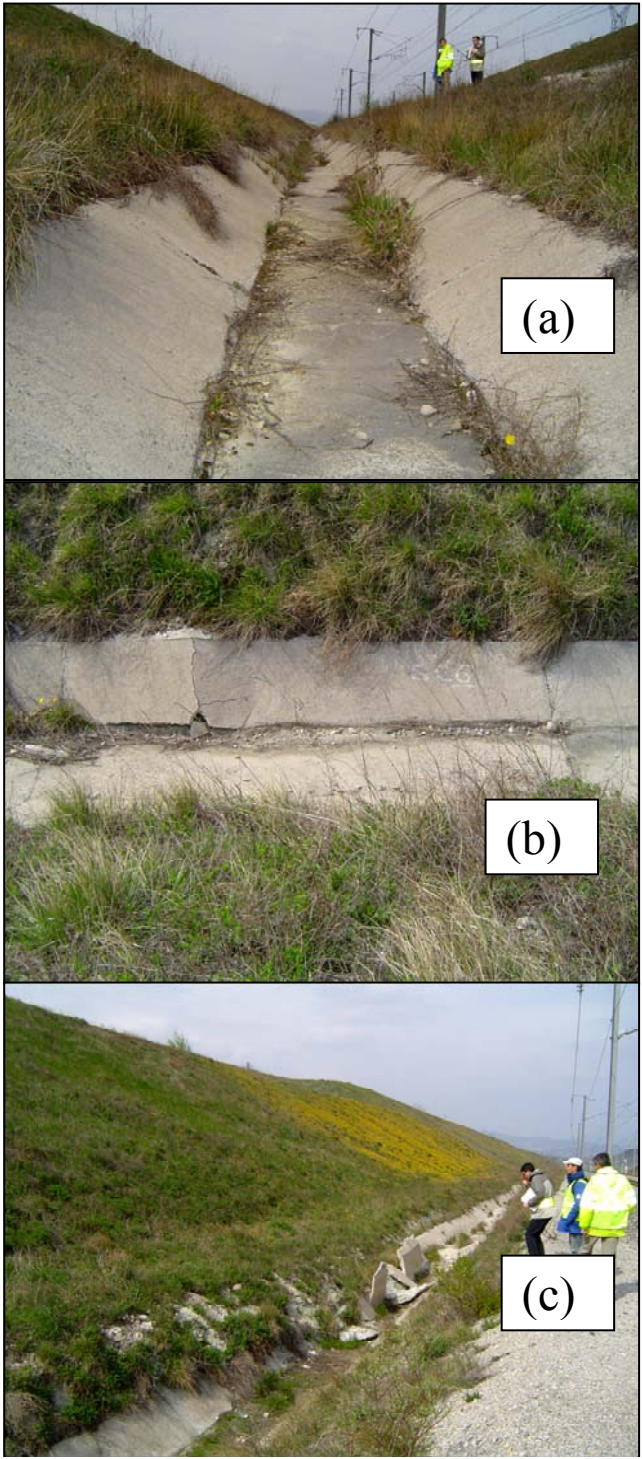

**Figure 3. Damages observed at the site of Chabrillan in April 2005 (pictures provided by SNCF).**

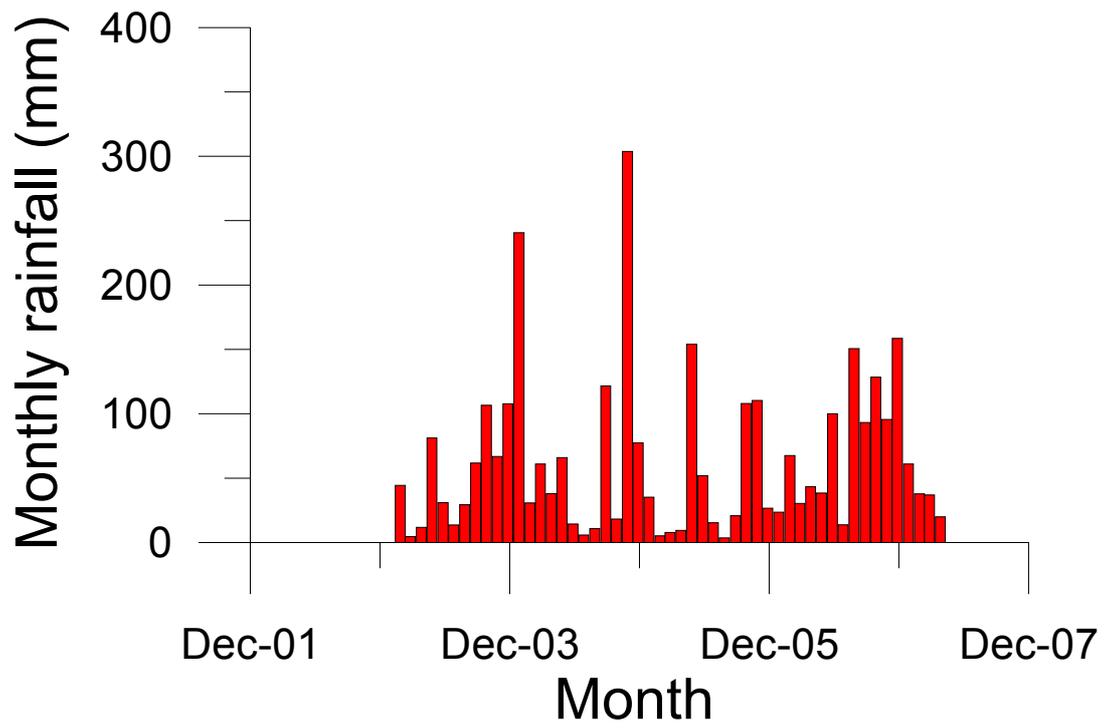

**Figure 4.** Monthly rainfall at Montélimar, 30 km in south-west of Chabrillan. (Data from http://meteociel.fr).

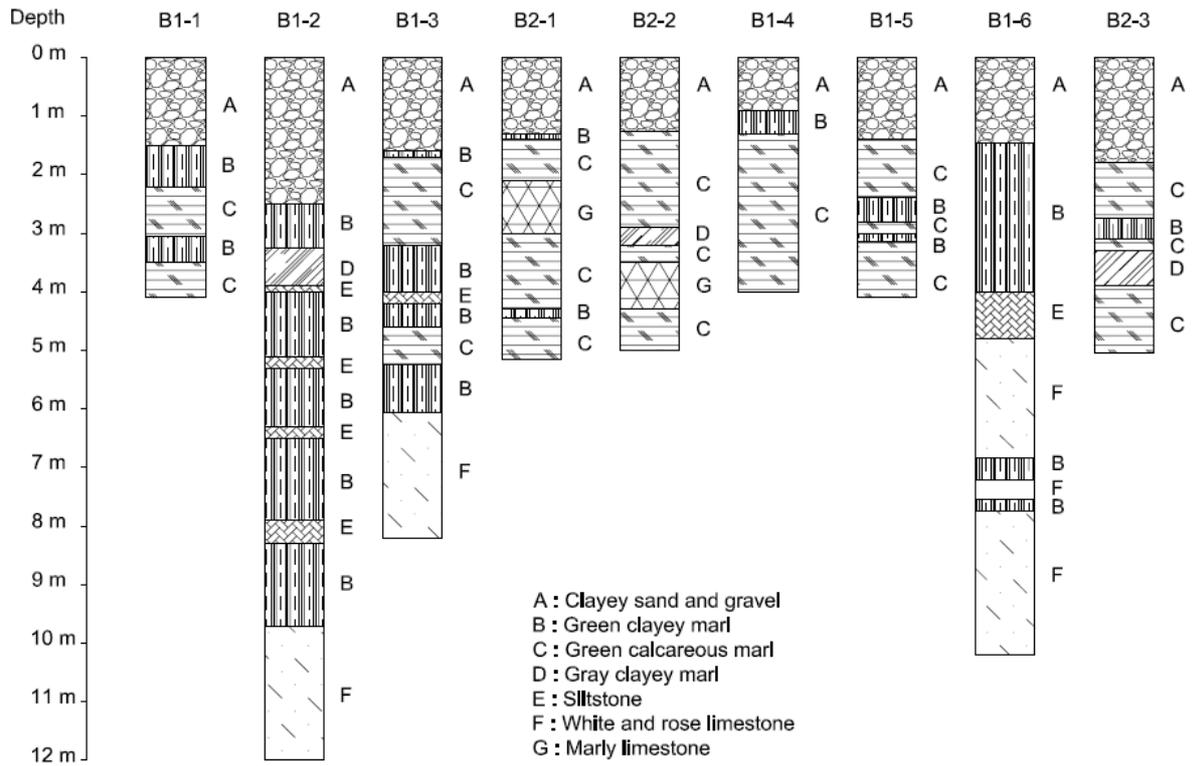

**Figure 5. Profiles of the boreholes. (Modified from Fondasol 2002 and data provided by SNCF in 2007).**

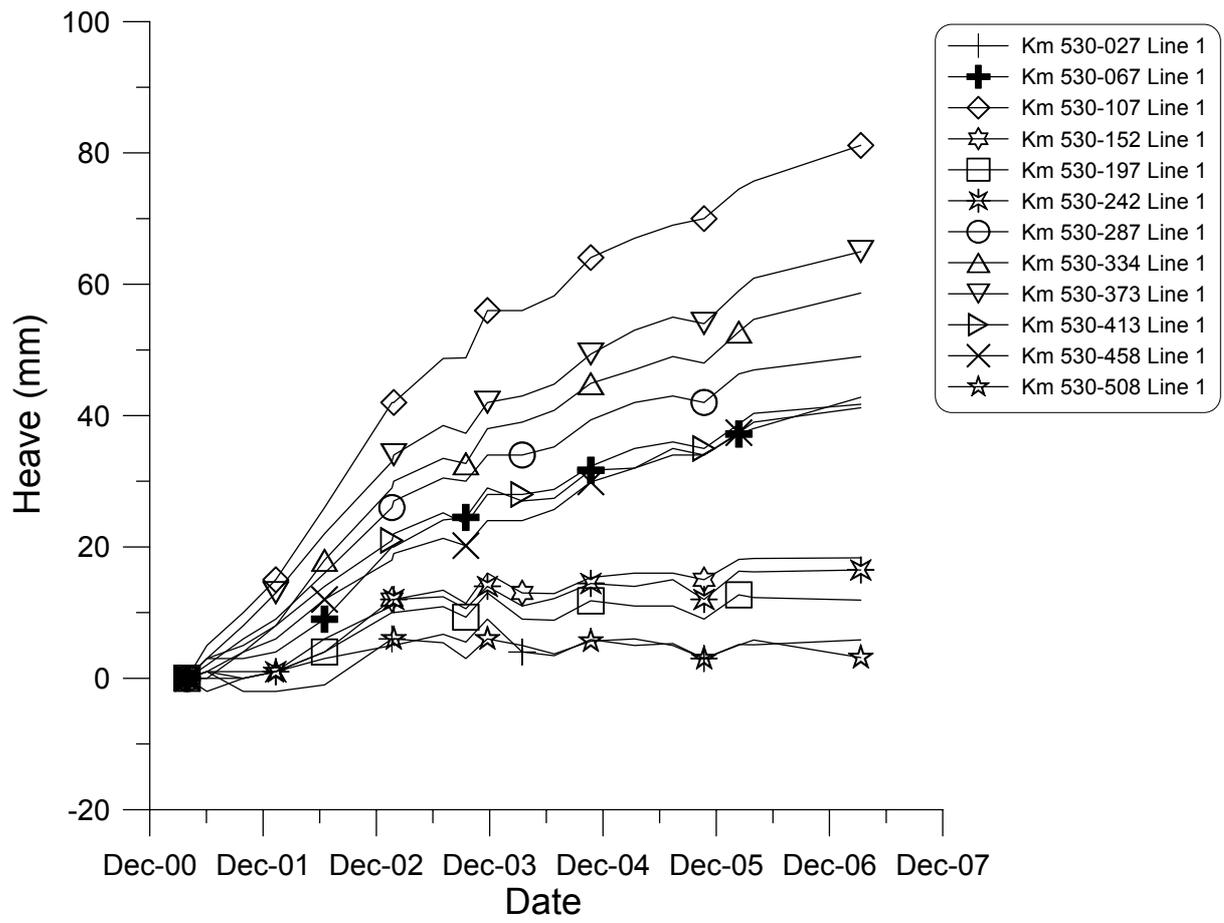

**Figure 6. Heave observed on the leg of the catenary poles of the Line 1 from April 2001. (after SNCF, 2007).**

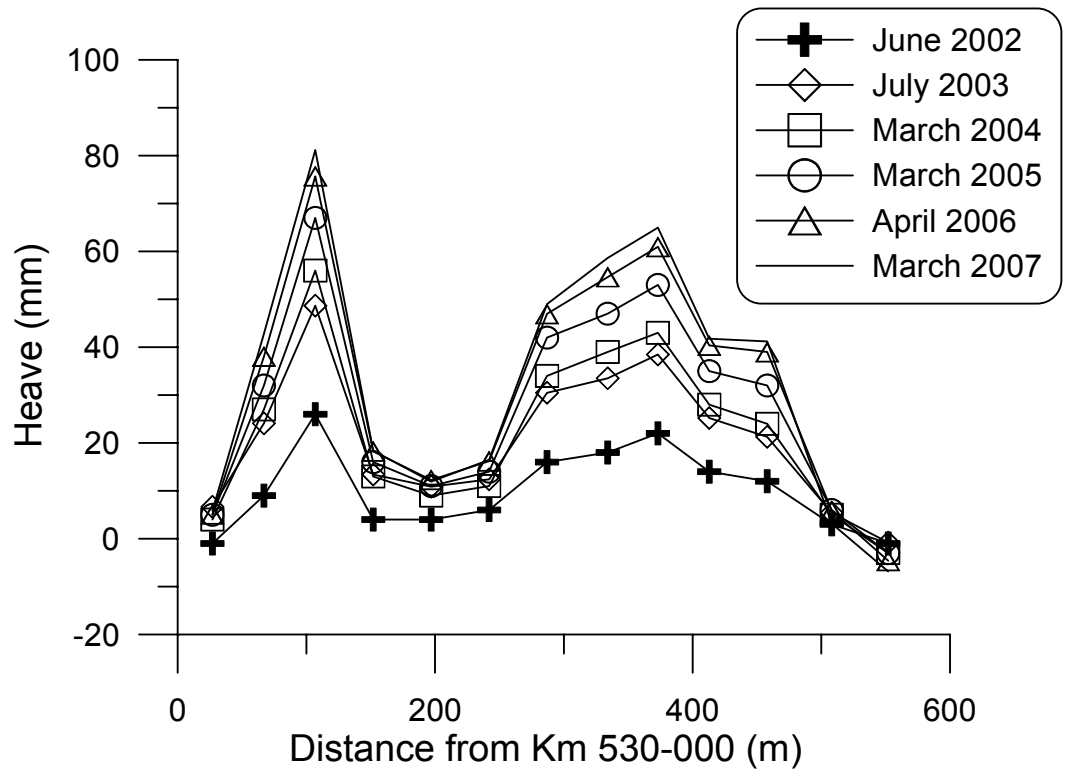

**Figure 7. Profile of heave observed on the leg of the catenary poles of the Line 1 from April 2001. (after SNCF, 2007).**

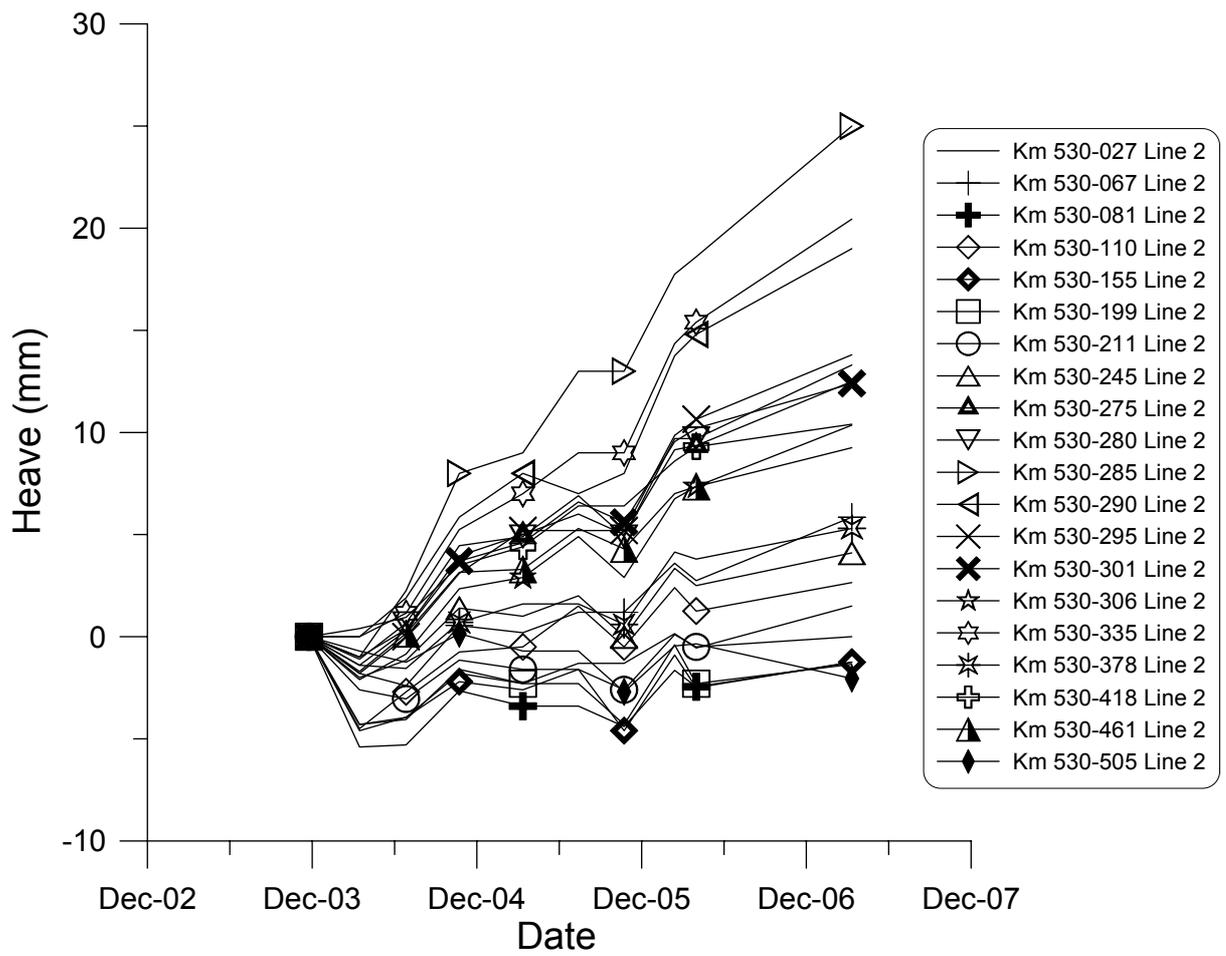

**Figure 8. Heave observed on the leg of the catenary poles of the Line 2 from November 2003. (after SNCF, 2007).**

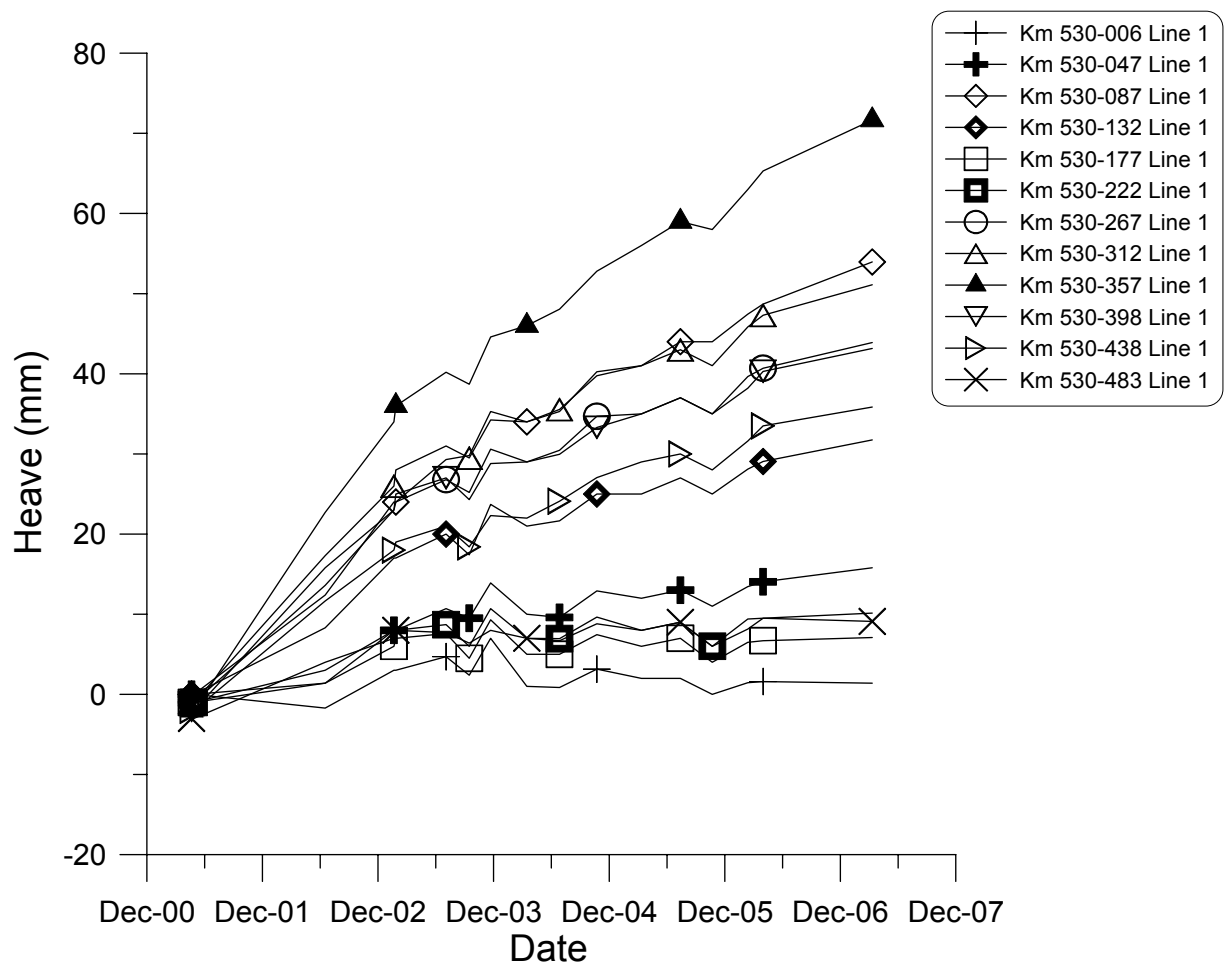

**Figure 9. Heave observed on the concrete railway sleepers of the Line 1 from April 2001. (after SNCF, 2007).**

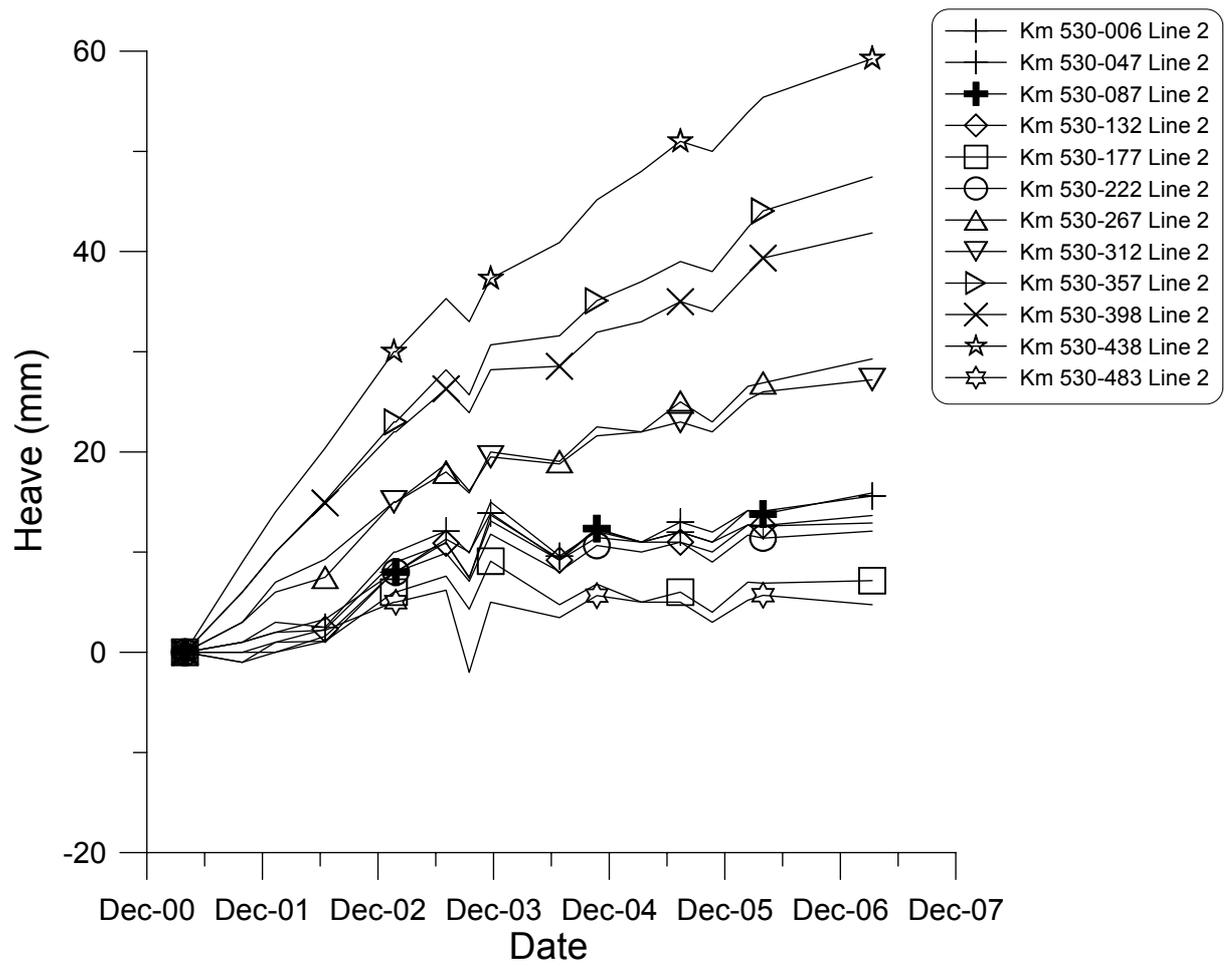

**Figure 10. Heave observed on the concrete railway sleepers of the Line 2 from April 2001. (after SNCF, 2007).**

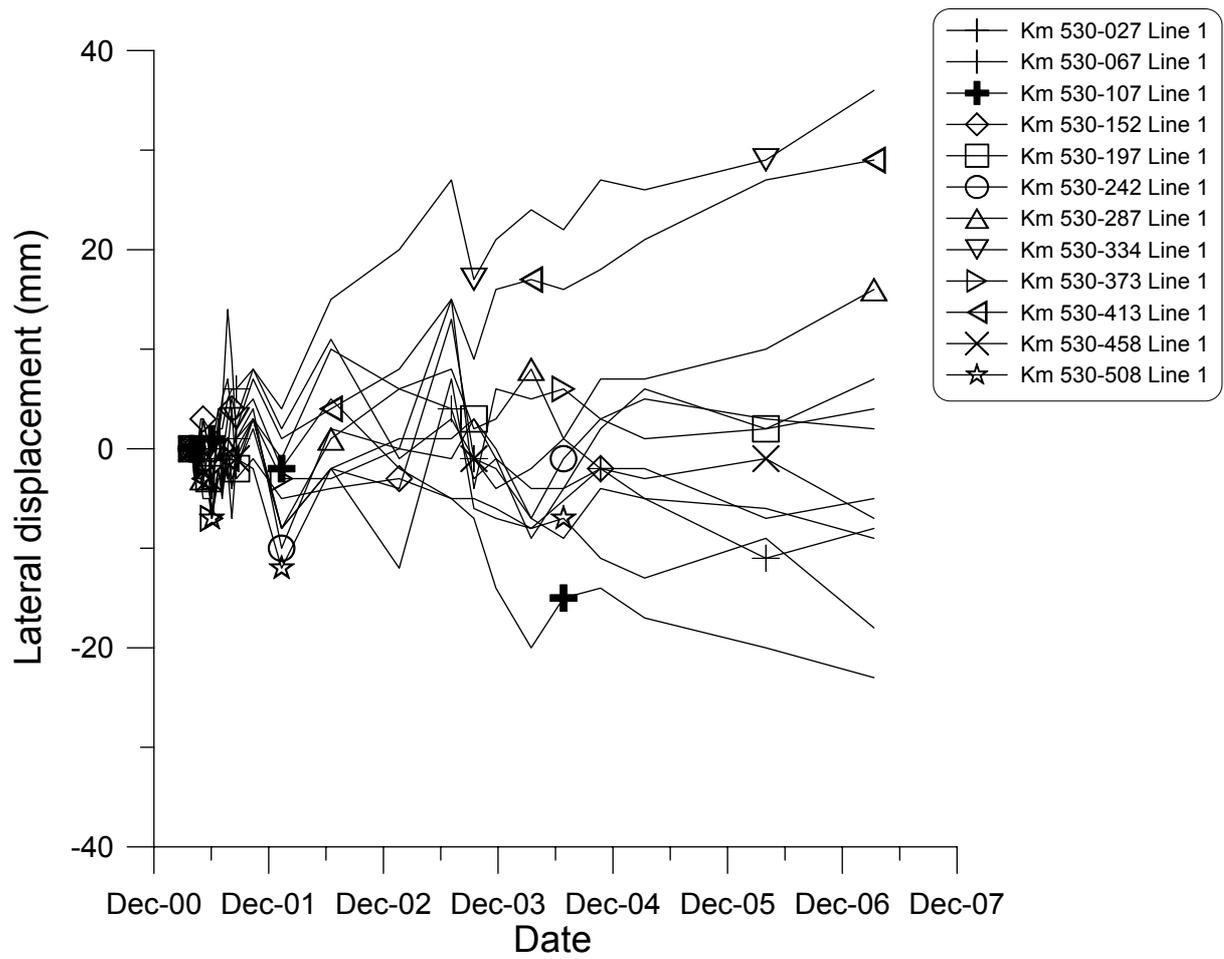

**Figure 11. Lateral displacement observed on the concrete railway sleepers of the Line 1 from April 2001. (after SNCF, 2007).**

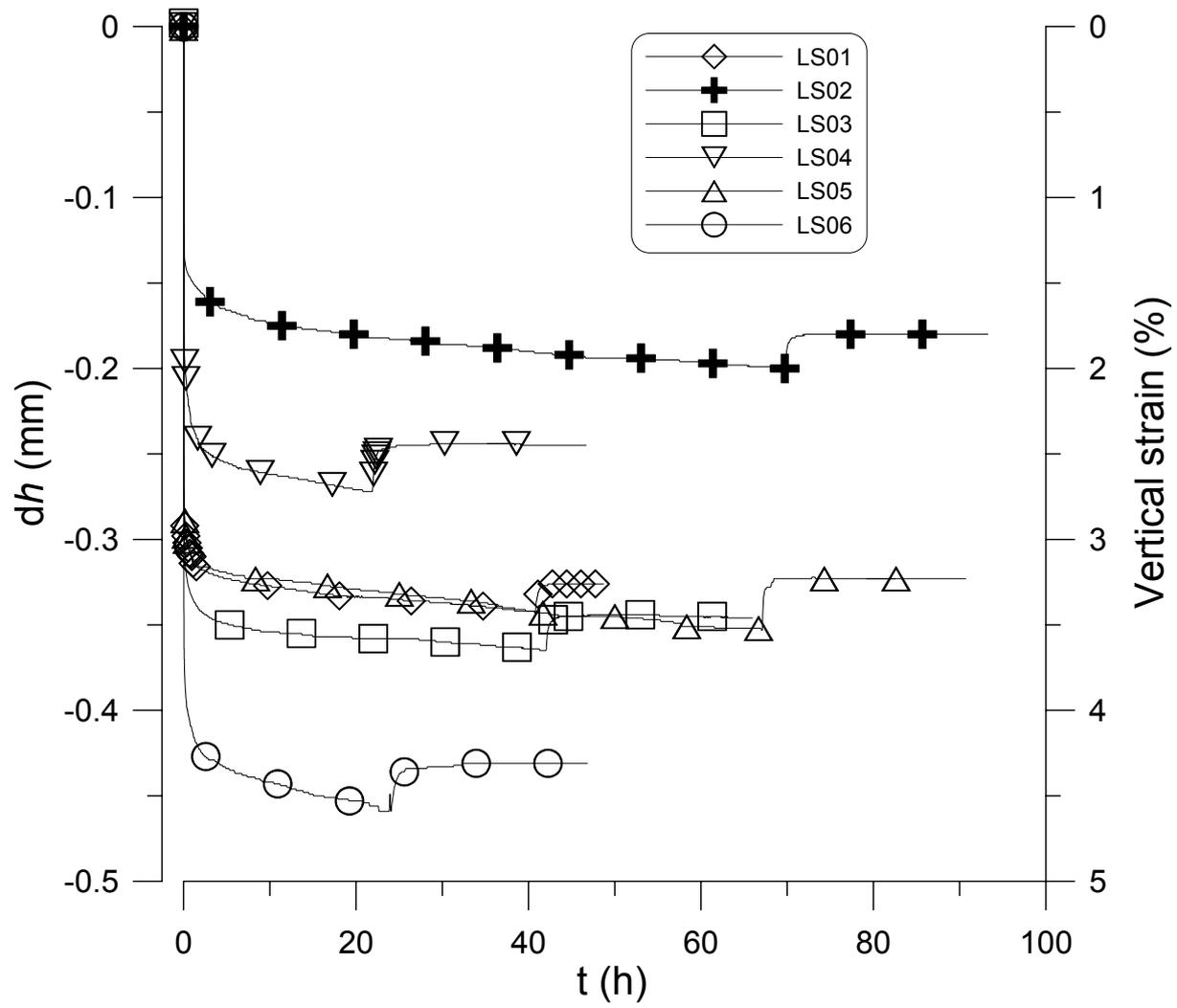

**Figure 12. Changes in height and vertical strains versus time for the samples tested following the loaded-swell method.**

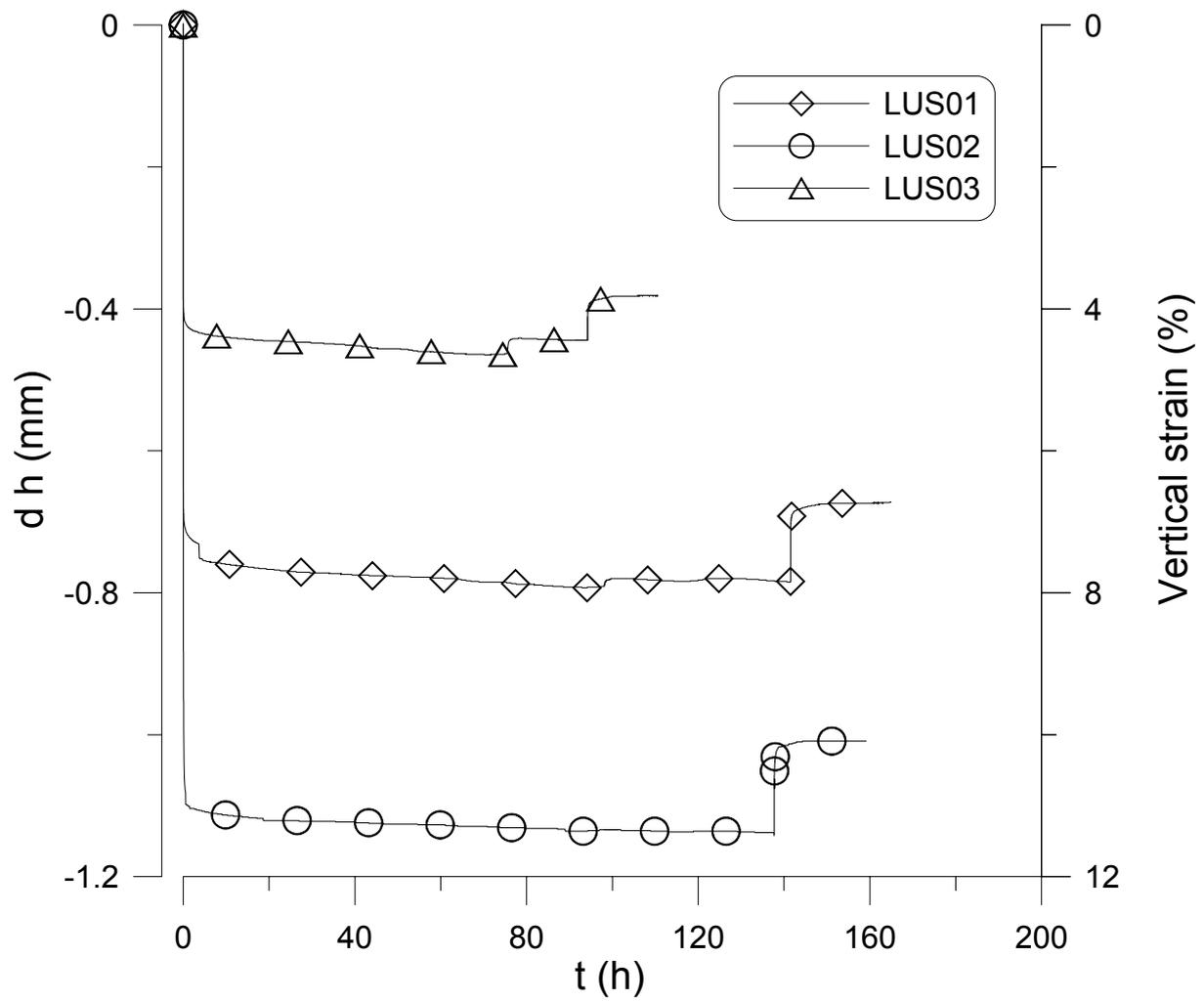

**Figure 13. Changes in height and vertical strains versus time for the samples tested following the loaded-unloaded-swell method.**

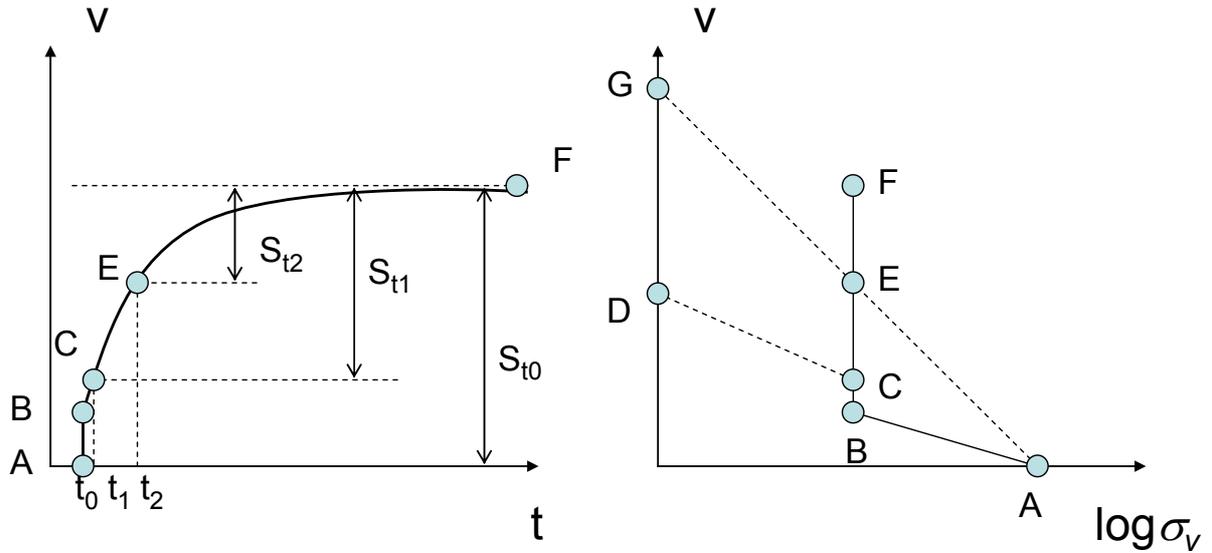

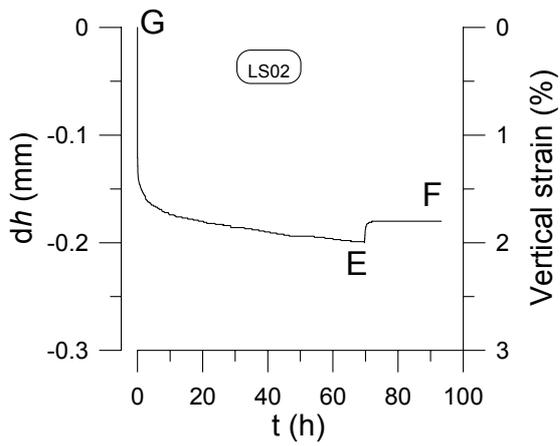
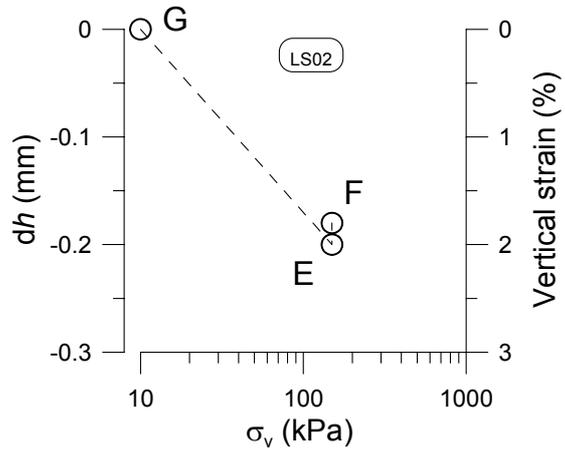

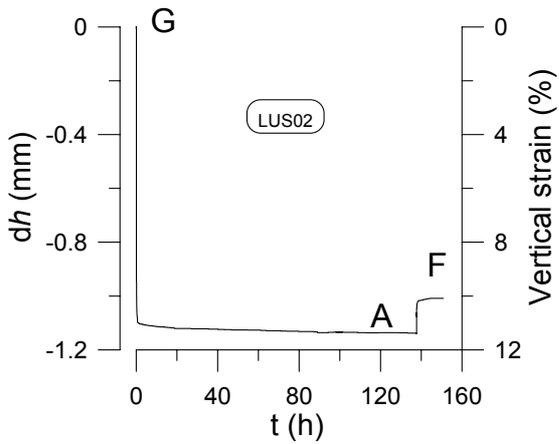
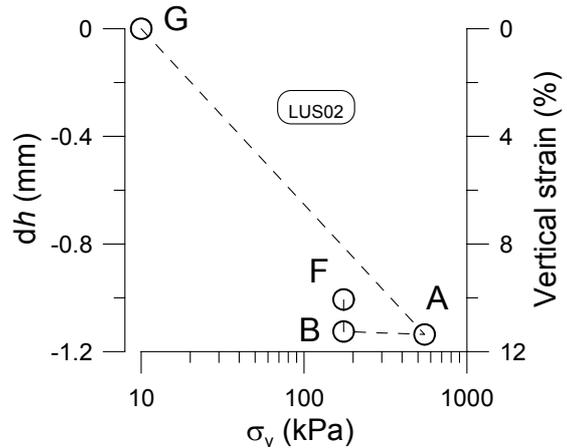

**Figure 14. Schematic view of the swelling mechanism (a, b) and the experimental results of test LS02 (load-swell method, c,d) and test LUS02 (load-unload-swell method).**

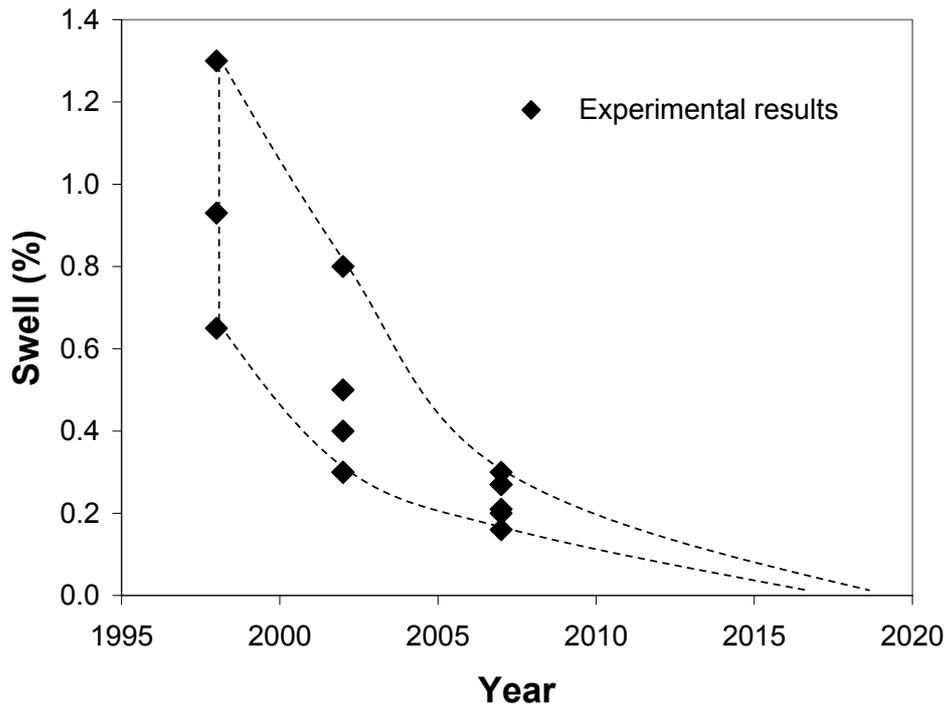

**Figure 15. Changes of swell potential versus time.**

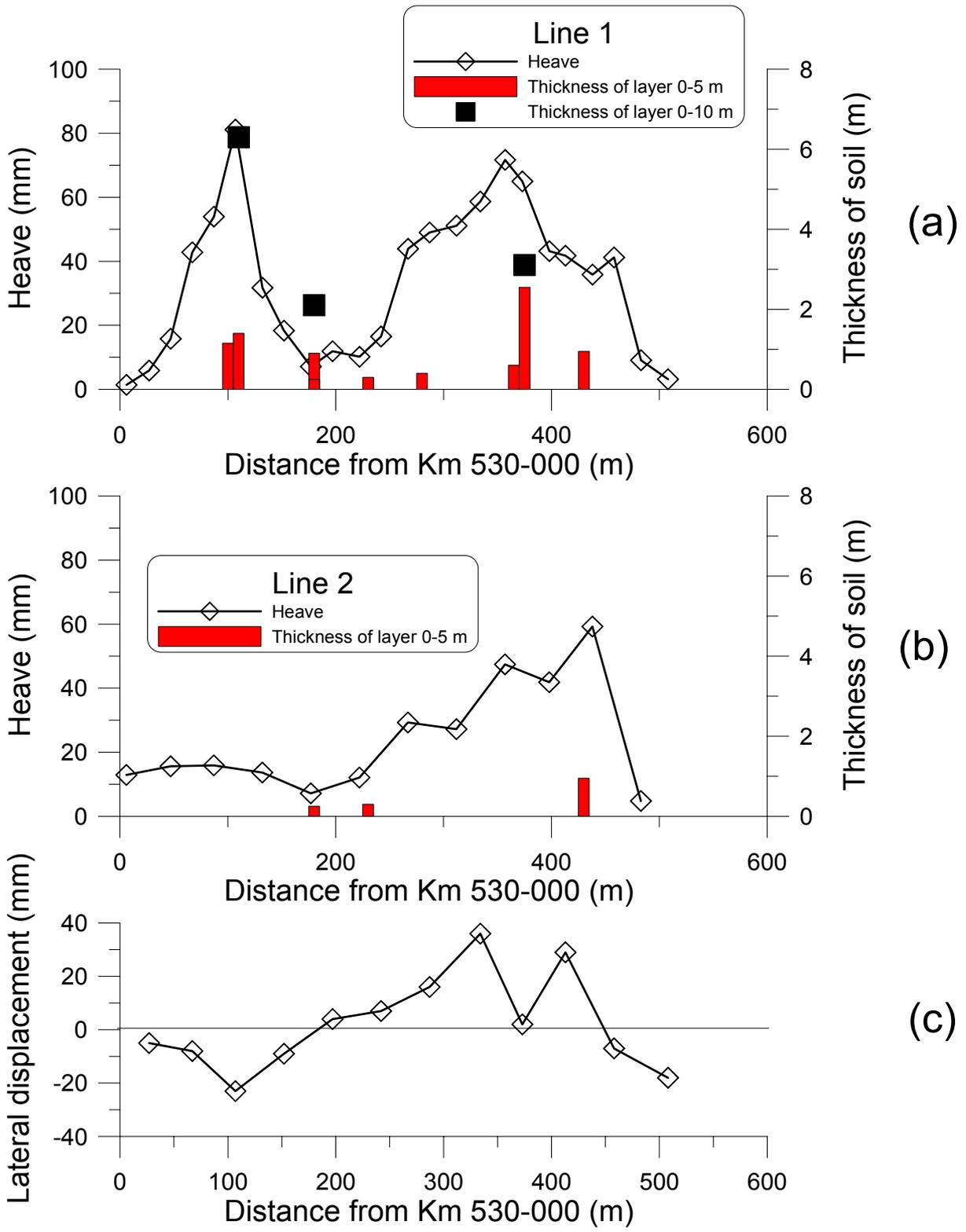

**Figure 16.** Heave at Line 1 (a), Line 2 (b) and the lateral displacement perpendicular to the line (c) between April 2001 and March 2007. The total thickness of the clayey marl layers was calculated from the boreholes profiles.